# Few-Layered Graphyne Growth by an On-Surface Coupling Reaction via Alkynyl Vapour Deposition

Sohyeon Seo,[1,†] Jungsue Choi,[1,†] Soo Min Cho,[1] Seungeun Lee,[1] and Hyoyoung Lee[1,2*]

[1]Department of Chemistry, Sungkyunkwan University (SKKU), Suwon 16419, Republic of Korea, [2]Centre for Integrated Nanostructure Physics (CINAP), Institute of Basic Science (IBS), Suwon 16419, Republic of Korea

[†]These authors contributed equally to work.

**Abstract**

A graphyne (GY) family composed of the triple (sp) and double ($sp^2$) bonds-hybridized carbon atoms is a promising allotrope of carbon for the development of nanoscale electronic devices. Unlike graphene, the GY family containing carbon sp bonds remains an unsolved problem in efficient two-dimensional (2D) growth due to monomer instability and side reactions. Herein, we synhesize GY into a single 2D layer by chemical vapor deposition (CVD) at low temperatures preventing unexpected reactions, using 1,3,5-tribromo-2,4,6-triethynylbenzene (TBTEB). The CVD-grown GY via a surface-confined coupling between TBTEB monomers on a copper surface can be rid of either one-dimensional (1D) or three-dimensional (3D) growths via side reactions. The CVD-grown GY exhibits a hexagonal lattice structure containing a single sp-carbon bond in a link between two adjacent hexagons, consisting with the structure of γ-GY. This work opens a new avenue for synthesizing large-scale and single-crystalline GY that can be capable of versatile applications with theoretically expected properties.

**Introduction**

Two-dimensional (2D) carbon allotropes such as graphene (composed of highly conjugated $sp^2$-carbon atoms) and graphyne (composed of hybridized sp- and $sp^2$-carbon atoms) have received scientific attention as promising candidates for future electronic, photonic, energy, catalytic applications.[1] For the past few decades, graphene that has a naturally missing bandgap have been developed toward its family (e.g., graphene quantum dots and graphene nanoribbons) possessing a bandgap as a semiconductor.[2] On the other hand, graphynes (GY) including graphdiynes (GDYs) are non-natural carbon allotropes that can be artificially designed to possess a natural bandgap. However, graphynes have been believed as theoretical materials for a few decades since the first structural model of graphyne was proposed.[3] Eventually, GDYs among the graphynes have been synthesized as results of many research efforts.[4-6] Up-to-date, the GDYs have been synthesized by various methods from on-surface reactions[7,8] to in-solution reactions,[9,10] and applied to several applications using their intrinsic pores[6,11,12]. Unfortunately, however, most of the synthesized GDYs are multi-layered that cannot unveil intrinsic properties. The electrical properties of few-layered GDY (e.g., γ-GDY) grown on other 2D materials such as h-BN and graphene exhibited conducting behaviours,[13,14] but which still remains unanswered issues for the intrinsic properties of GDY itself. On the one hand, GY (e.g., γ-GY) that has a bandgap like γ-GDY has been synthesized via cross-coupling between $sp^2$-carbon and sp-carbon by a mechanochemical reaction recently,[15,16] but single- or few-layered GY like GDY was hardly obtained. Thus, exploring new synthetic methods toward obtaining single- or few-layered GYs including GDYs are required for their promising challenges in scientific and technological developments.

Chemical vapour deposition (CVD) method is well-known to enable the large-scale growth of high-quality and 2D single-crystalline graphene. According to the previous report, however, CVD-grown GDY using hexaethynylbenzene (HEB) or 1,3,5-triethynylbenzene (TEB) at high temperatures synthesized amorphous carbon films, indicating that side reactions occurred against a Glaser-coupling reaction between terminal ethynyls.[4] Unlike graphene, the backbone (sp- and $sp^2$-hybridized carbon backbone) of precursor monomers should be maintained in GYs. Thus, in this work, to synthesize GY composed of a single sp-carbon bond in a link between two adjacent hexagons, 1,3,5-tribromo-2,4,6-triethynylbenzene (TBTEB) (Supporting Information, Scheme S1) was synthesized to promote a cross-coupling reaction between terminal ethynyls and bromines instead of a homo-coupling reaction between terminal bromines (Scheme 1a). Surface-assembled TBTEB molecules can follow a cross-coupling (e.g., Castro-Stephens coupling[17]) on Cu foil. CVD growth of GY is performed at low-temperature conditions to prevent decomposition of the monomers and side reactions. Moreover, a two-step process of (1) surface assembly and (2) surface reaction under different temperature conditions can prefer the cross-coupling reaction to be competitive advantage compared with the homo-coupling reaction. Raman shifts and x-ray photoelectron spectra of CVD-grown GY sheets revealed the chemical and structural information of γ-GY. Selected area electron diffraction (SAED) patterns and x-ray diffraction spectra of CVD-grown GY confirmed well-crystalline γ-GY. Our approach provides new insights into synthesizing large-scale and single-crystalline GY.

**Results and discussion**

CVD growths of GY have been conducted by programmed temperature control in two separate zones of a vacuum tube under a stream of Ar carrier gas (Scheme 1a and Supporting Information, Figure S1a,b). Monomers were evaporated in a precursor zone at 40 °C, transferred in the carrier gas stream to onto Cu foil in a growth zone, and assembled on Cu foil at room temperature or 60−80 °C. The assembled monomers on Cu foil were grown into GY via a cross-coupling reaction at 60−80 °C (Scheme 1b, Supporting Information Scheme S1c). The synthesized film was transferred onto a silicon oxide substrate. As shown in an optical microscope (OM) image (Supporting Information, Figure S2a) and a scanning electron microscope (SEM) image (Figure 1a), a large-scale film was obtained. Chemical information characteristics of the synthesized film at the low temperature were explored by Raman spectroscopy (Figure 1b), x-ray photoelectron spectroscopy (XPS) (Figure 1c,d), and Fourier-transform infrared (FT-IR) spectroscopy (Supporting Information, Figure S2b). In the Raman spectrum, the G and 2D band involving $sp^2$-conjugated carbons were observed at 1444 and 2750 $cm^{-1}$, respectively. The G band is originated from the stretching vibration of $sp^2$-hybridized carbon atoms in benzene rings. The observed G band in this study was negatively shifted from that of graphene comprised of $sp^2$-hybridized carbon atoms. This might be the influence of ethynyls linked to benzene rings. The D band around 1300 $cm^{-1}$ was attributed to the defects and disordered structures. In particular, a fingerprint peak of sp-carbon from conjugated ethynyl links was observed at 2187 $cm^{-1}$ in the Raman spectrum[18] and at 2342 $cm^{-1}$ in the FT-IR spectrum[5]. Moreover, the XPS spectrum of the synthesized film did not show a peak of Br 3d corresponding to TBTEB. Although oxidation peaks were observed, the deconvoluted C 1s spectrum of XPS obviously verified $sp^2$ (C=C) at

284.3 eV and sp (C-C) at 284.9 eV.[19] The area ratio of sp/sp$^2$ was approximately 1, verifying that sp$^2$-hybridized carbon atoms in benzene rings were linked to each other via one sp-hybridized carbon atom. This consists with the unique structural information of γ-GY.[20,21] On the other hand, the sp$^3$-hybridized carbon was not observed in the C 1s spectrum. The strong oxygen peak in (c) and oxidized carbon peaks in (d) could be attributed to the oxygen on the SiO$_2$ substrate and the surface oxidation of the film during either growing or transferring processes.

To confirm the crystal structure of CVD-grown γ-GY, x-ray diffraction (XRD) spectroscopy was performed. Theoretically, γ-GY is the hexagonal system that has the hexagon patterns ascribed to (110), (221), (220), and (222) lattice planes.[15] The XRD patterns of the CVD-grown γ-GY film transferred on a SiO$_2$ substrate obviously showed the peak at 34.2 ° (d = 0.25 nm) corresponding to the theoretical interplanar spacing of the (222) lattice plane (Figure 2a), while the (220) lattice plane corresponding to 25.4 ° (d = 0.35 nm) was hardly observed. The XRD patterns of TBTEB disappeared in the XRD spectrum of the γ-GY film. Moreover, the morphology and crystal structure of the γ-GY film was explored by transmittance electron microscopy (TEM). The CVD-grown γ-GY/Cu foil was transferred onto a TEM grid. Figure 2b shows a TEM image of few-layered γ-GY sheets grown at 60 °C after assembly at room temperature with vaporizing at 40 °C. The high-resolution TEM (HRTEM) image clearly showed the crystalized morphology in the γ-GY sheet (Figure 2c). The fast Fourier transform (FFT) patterns and the selected area electron diffraction (SAED) patterns confirmed the theoretical lattice constant of the (222) lattice plane of γ-GY (Figure 2d), which is identical to the XRD result. In fact that the spots in the SAED patterns showed one hexagon of the (222) lattice planes only, the γ-GY sheets were orderly stacked by the way of ABC trilayer stacking as expected in GDY (Supporting Information, Figure S3a).[22] In addition, the distance between spots in the HRTEM image was measured by 0.28 nm (in the zoomed-in image as the upper inset in Figure 2c). Thus, the ABC trilayer can be configured on the image as shown in Figure 2e. Moreover, the titled hexagon spots in the SAED pattern were also observed, indicating the bilayer or multilayer growths of γ-GY, where the upper γ-GY sheet was also grown on the lower γ-GY with slightly tilted angles (i.e., titled bilayer stacking) (Supporting Information, Figure S3b-d). In this case, the spots corresponding to the (110) lattice planes (d = 0.69 nm) were observed due to the titled bilayer stacking.[22]

Furthermore, as the assembly temperature increased to the growth temperature (60−80 °C), γ-GY was grown into thicker sheets, as shown in the TEM images in Figure 3a,d. The HRTEM images of both γ-GY sheets grown at 60 and 80 °C also showed obvious crystalline morphologies (Figure 3b,e) clearly patterned by a single crystal, as shown in FFT images. From the SAED pattern analyses, two-type patterns were obtained as observed in the γ-GY grown at 60 °C after assembly at room temperature; (1) the SAED pattern with a clear single hexagon corresponding to the (222) lattice planes only confirmed that γ-GY was mainly grown into ABC trilayer-stacked sheets at both temperatures (Figure 3c,f). (2) At 60 °C, the SAED patterns of the γ-GY sheet showed slightly eclipsed hexagons and confirmed that the γ-GY sheets were grown into multilayers by slightly tilted bilayer growths (Supporting Information, Figure S4a,b). At 80 °C, the γ-GY sheet exhibited more eclipsed hexagon patterns in the FFT image than that grown at 60 °C, indicating that the upper layers were grown on the lower layers at 80 °C with large titled angles (Supporting Information, Figure S4c,d). Consequently, in cases of high-temperature assemblies, the layered growths of γ-GY sheets occurred by the titled bilayer stacking.

**Experimental section**

**Materials**: 1,3,5-tribromo-2,4,6-triethynylbenzene (TBTEB) was synthesized according to the previous report.[23] The synthesized TBTEB monomer was stored under −20 °C.

**GY synthesis**: 0.25 micron-thick Cu foil (Alfa aesar) was firstly prepared by chemical cleaning using acetic acid (Alfa aesar), washed with deionized (DI) water thoroughly, and dried with nitrogen stream. The cleaned Cu foil was transferred into a CVD chamber comprised of two quartz tubes and thermally annealed at 890 °C for h under a vacuum. After the temperature was cooled down to room temperature, a vial containing TBTEB monomers was placed in heat zone 1 and the Cu foil was placed in heat zone 2. GY films synthesized on Cu foil were transferred onto cleaned $SiO_2$ substrates immediately according to the previous report for the transferring method of CVD-grown graphene. The $SiO_2$ substrates were pre-cleaned with piranha solution, rinsed with DI water several times, dried by blowing $N_2$, and then treated with oxygen plasma.

**Characterization**: CVD-grown GY sheets on $SiO_2$ were used for characterization by using XPS (Thermo VG, Microtech ESCA), Raman (UHTS300, WITec), XRD (Ultima IV, Rigaku), SEM (SEM, JSM-7100F, JEOL), FT-IR (Tensor27, Bruker), and HRTEM (JEM-2100F, JEOL). All measurements were performed at room temperature. XPS measurements were performed with a monochromatic Al-Kα X-ray source operated at 100 W. The binding energies obtained in the XPS analysis were corrected for specimen charging by referencing to the C 1s line at 288.4 eV. HRTEM was performed at 200 keV.

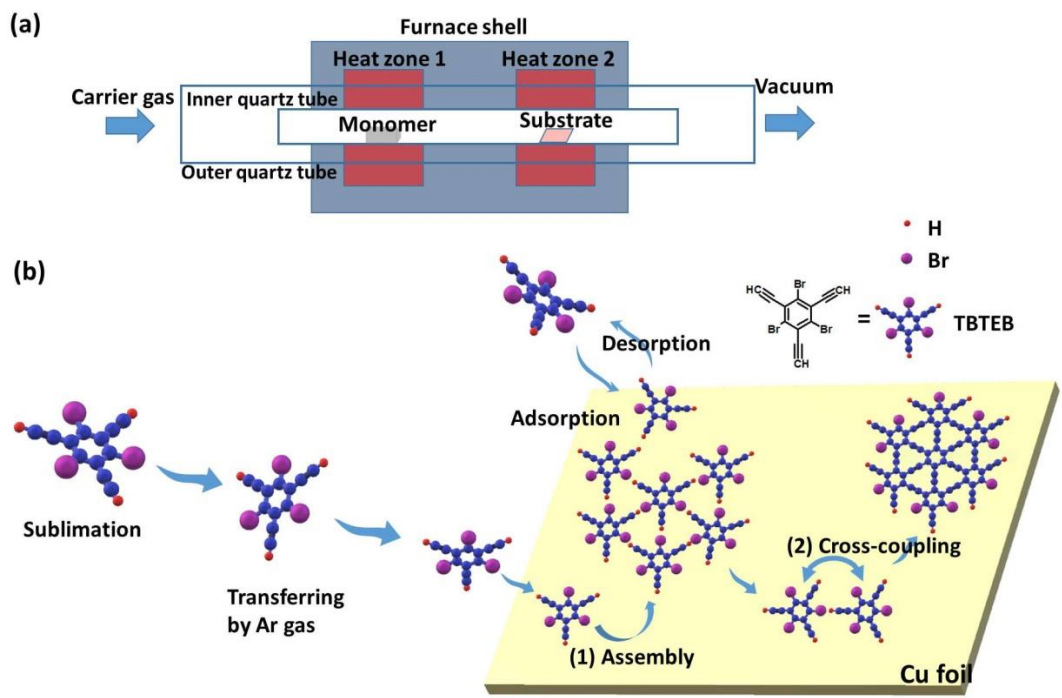

**Scheme 1**. Illustrations of (a) a vacuum furnace for chemical vapor deposition (CVD) and (b) a progress of surface-assisted cross-coupling reactions of HBTEB in a CVD chamber.

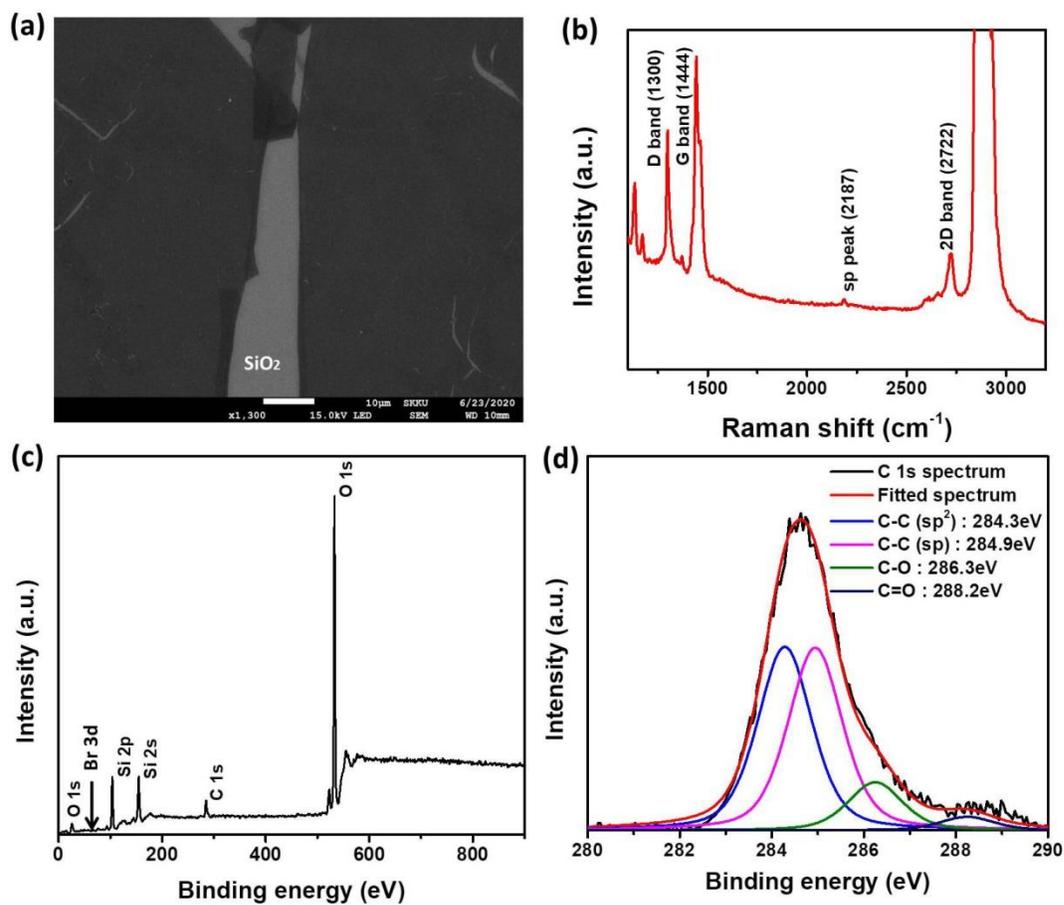

**Figure 1**. Chemical information of CVD-grown graphyne (GY) film. (a) Scanning electron microscope (SEM) image. (b) Raman spectrum collected with 532 nm excitation. (c) and (d) Wide-scan and C 1s spectrum of x-ray photoelectron spectroscopy (XPS), respectively.

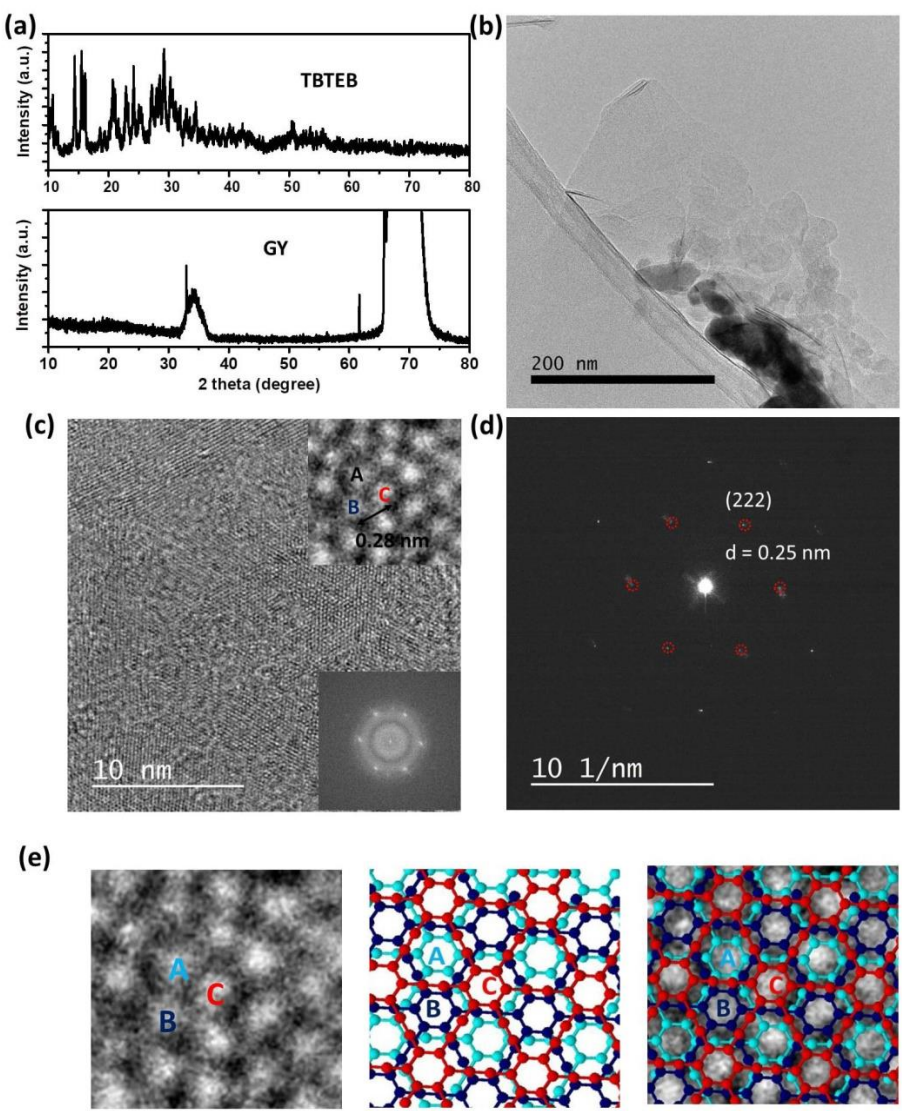

**Figure 2**. (a) X-ray diffraction (XRD) spectra of TBTEB and CVD-grown GY. (b) Transmittance electron microscope (TEM) image of GY grown at 60 °C after sublimation at 40 °C and assembly at room temperature. (c) High-resolution TEM (HRTEM) image of a single layer GY sheet. Upper inset presents the zoomed-in HRTEM image and the positions of ABC trilayer stacking. Lower inset image presents the fast Fourier transform (FFT) pattern corresponding (c). (d) Selected area electron diffraction (SAED) pattern of a single layer GY

sheet. The red dotted circles mark a hexagonal pattern representing the (222) lattice planes. (e) The zoomed-in HRTEM image in (c), proposed ABC trilayer stacking, and the overlapped image.

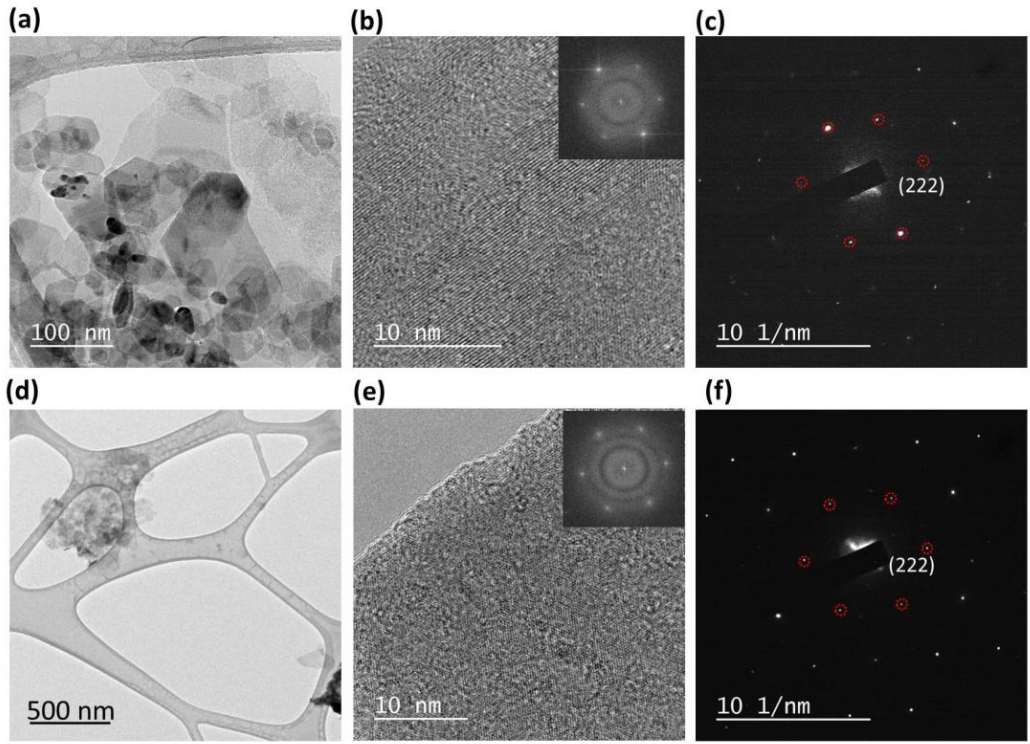

**Figure 3**. Temperature influences on CVD growths of GY. (a)-(c) Sublimation at 40 °C and assembly/growth at 60 °C. (a) TEM image of CVD-grown GY sheets, (b) the HRTEM image (inset: the correspondent FFT pattern) and (c) the SAED pattern of a trilayer-stacked GY sheet. The red dotted circles mark a hexagonal pattern representing the (222) lattice planes. (d)-(f) Sublimation at 40 °C and assembly/growth at 80 °C. (d) TEM image of CVD-grown GY sheets, (e) the HRTEM image (inset: the correspondent FFT pattern) and (f) the SAED pattern of a trilayer-stacked GY sheet. The red dotted circles mark a hexagonal pattern representing the (222) lattice planes.

# Supporting Information

Few-Layered Graphyne Growth by an On-Surface Coupling Reaction via Alkynyl Vapour Deposition

Sohyeon Seo,[1,†] Jungsue Choi,[1,†] Soo Min Cho,[1] Seungeun Lee,[1] and Hyoyoung Lee[1,2*]

[1]Department of Chemistry, Sungkyunkwan University (SKKU), Suwon 16419, Republic of Korea, [2]Centre for Integrated Nanostructure Physics (CINAP), Institute of Basic Science (IBS), Suwon 16419, Republic of Korea

[†]These authors contributed equally to work.

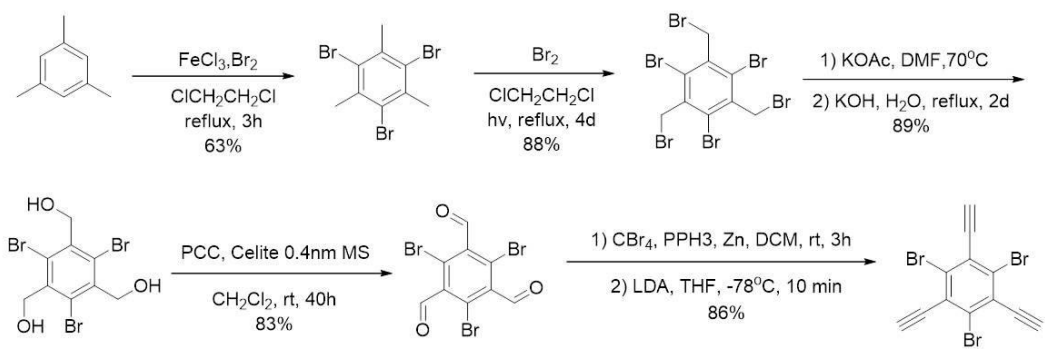

**Scheme S1**. Synthetic process of 1,3,5-tribromo-2,4,6-triethynylbenzene (TBTEB).

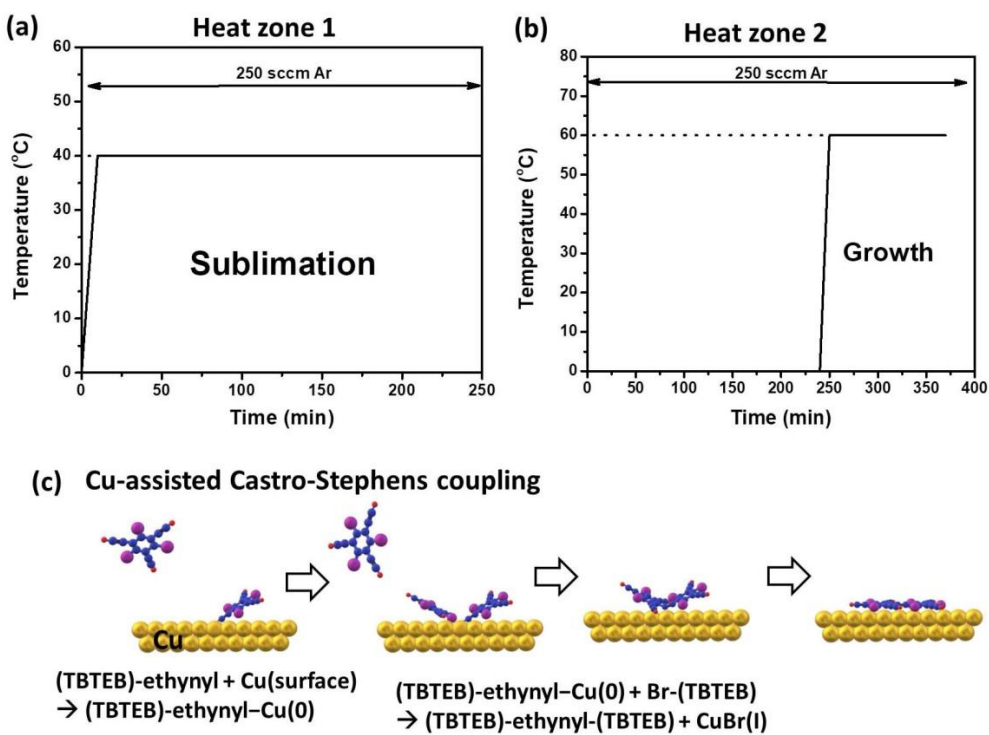

**Figure S1**. (a) and (b) Schematics of programmed temperature control in a vacuum furnace for chemical vapor deposition (CVD) growths of graphyne. (c) Illustrations of a Cu-assisted Castro-Stephens coupling reaction of TBTEB on Cu foil.

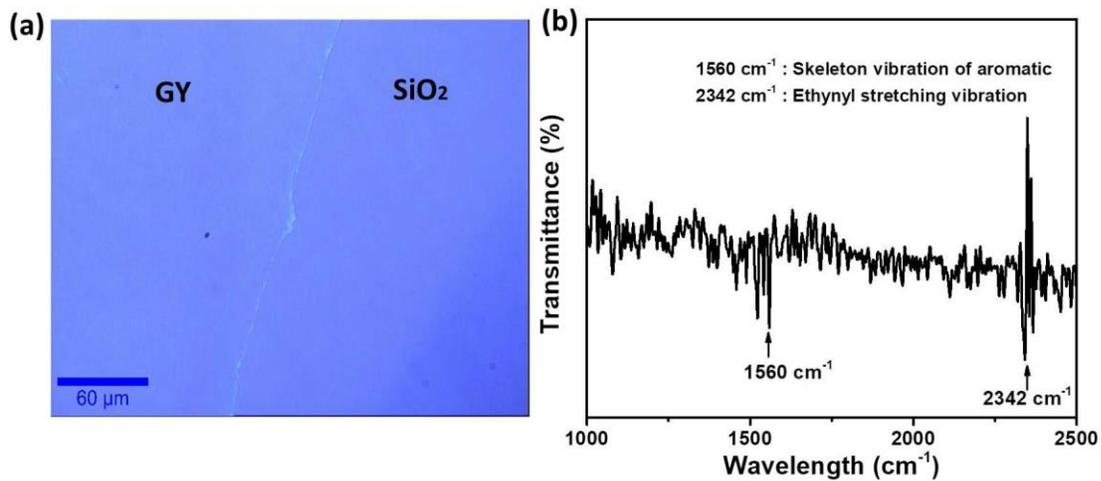

**Figure S2**. **(a)** Optical microscope image of CVD-grown GY after transferring on a SiO$_2$ substrate. **(b)** Fourier-transform infrared (FT-IR) spectrum of CVD-grown GY after transferring on a SiO$_2$ substrate.

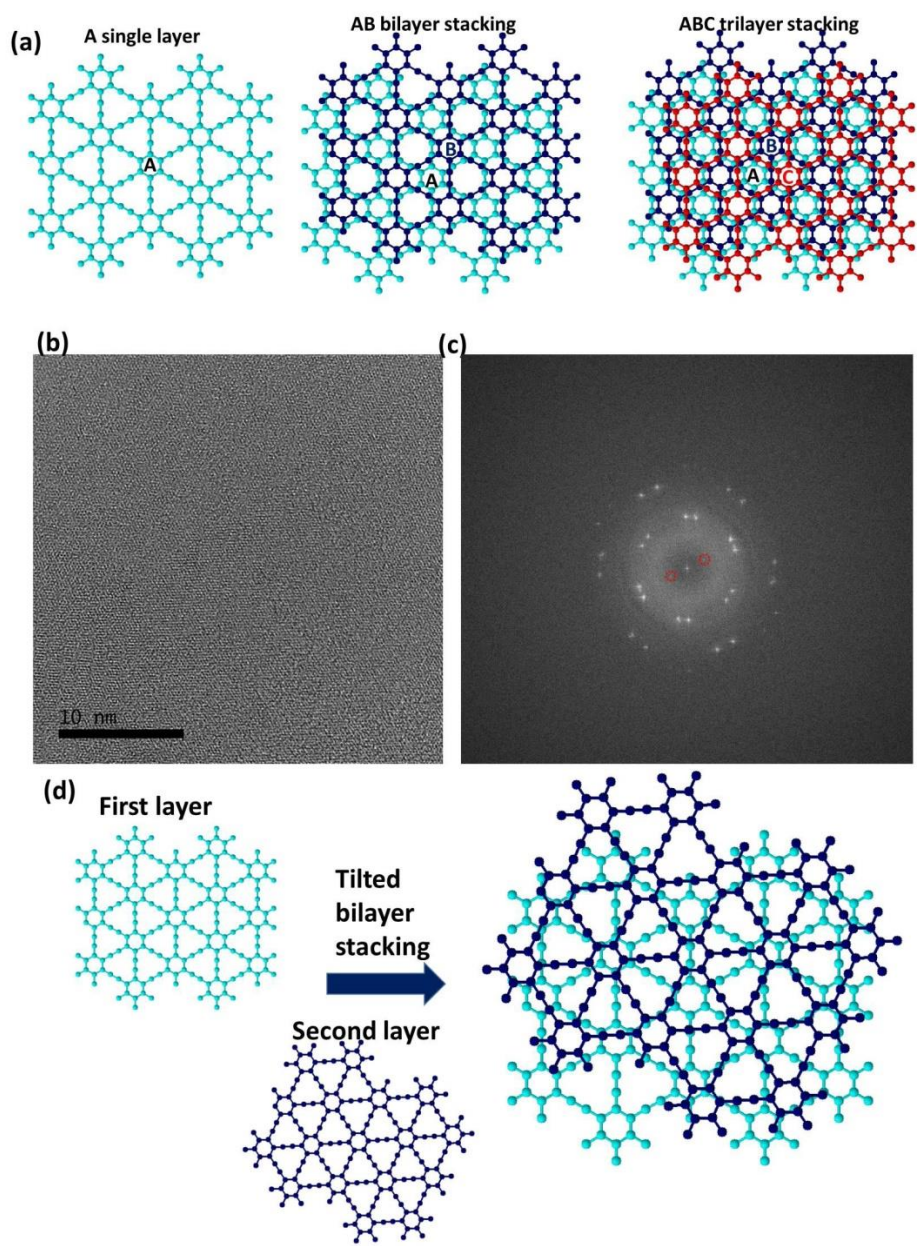

**Figure S3.** (a) Illustrations for one of possible ABC trilayer stackings of γ-GY. (b) High-resolution transmittance electron microscope (HRTEM) image and (c) fast Fourier transform (FFT) patterns of γ-GY indicating a multilayer growth. The red dotted circles in (b) mark the (110) lattice planes. (d) Illustrations of a bilayer γ-GY sheet formed by titled bilayer stacking.

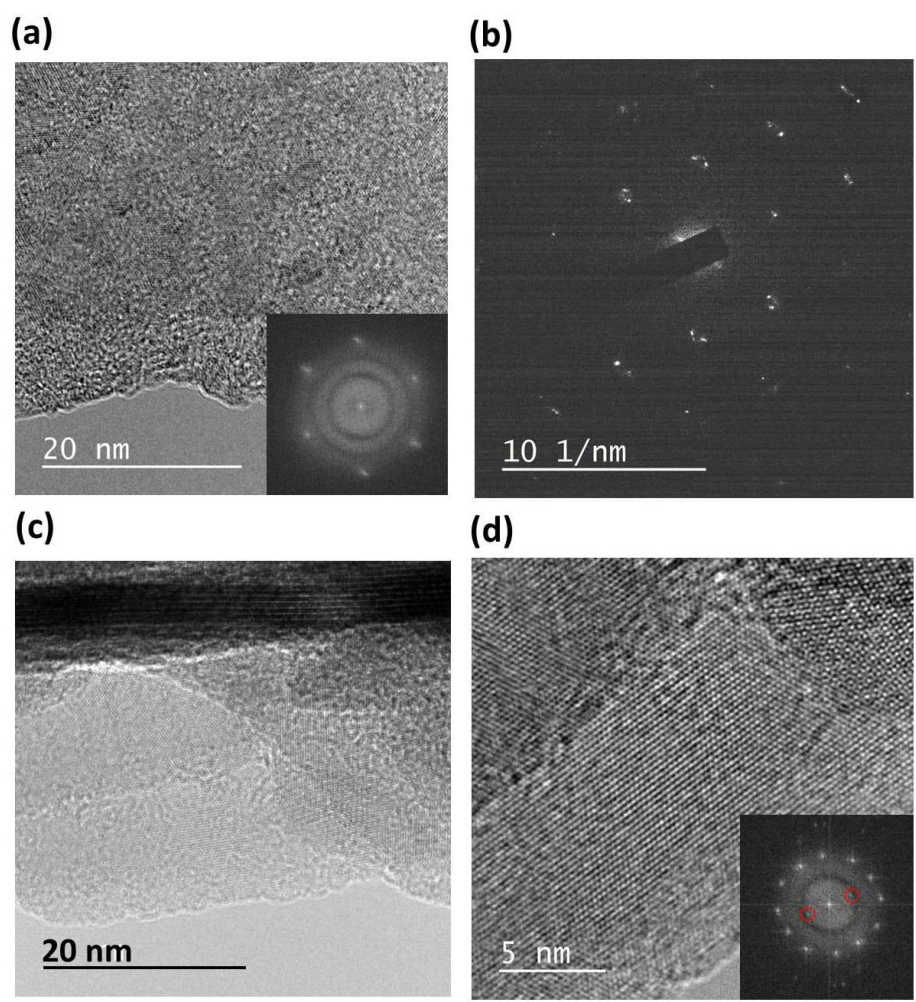

**Figure S4.** Temperature influences on CVD growths of GY. (a) and (b) Sublimation at 40 °C and assembly/growth at 60 °C. (a) HRTEM image of CVD-grown GY sheets (inset: the correspondent FFT pattern) and (b) the SAED pattern. (c) and (d) Sublimation at 40 °C and assembly/growth at 80 °C. (c) HRTEM image of CVD-grown GY sheets (inset: the correspondent FFT pattern) and (d) the SAED pattern. The red dotted circles mark the (221) lattice planes.